\documentclass{ws-mpla}

\usepackage[papersize={8.5in,11in}]{geometry}
\usepackage{verbatim}
\usepackage{hyperref}

\newcommand{\prd}{{\it Phys.~Rev.~D} }
\newcommand{\prl}{{\it Phys.~Rev.~Lett.} }

\newcommand{\harxold}[1]{\href{http://arxiv.org/abs/hep-lat/#1}{[
$\!\!$arXiv:hep-lat/#1]} }
\newcommand{\harxnew}[1]{\href{http://arxiv.org/abs/#1}{[arXiv:#1]} }

\newcommand{\MS}{\overline{MS}}

\newcommand{\python}{\texttt{{\normalsize P}{\footnotesize YTHON}}~}
\newcommand{\fortran}{\texttt{{\normalsize F}{\footnotesize ORTRAN}}~}
\newcommand{\vegas}{\texttt{{\normalsize V}{\footnotesize EGAS}}~}
\newcommand{\taylor}{\texttt{{\normalsize T}{\footnotesize aylUR}}~}
\newcommand{\hippy}{\texttt{{\normalsize H}{\footnotesize I}{\normalsize
PP}{\footnotesize Y}}~}
\newcommand{\hpsrc}{\texttt{{\normalsize HP}{\footnotesize SRC}}~}
\newcommand{\hippycomma}{\texttt{{\normalsize H}{\footnotesize
I}{\normalsize PP}{\footnotesize Y}},~}

\newcommand{\pastor}{\texttt{pastor}~}

\newcommand{\msbar}{\overline{MS}}

\begin{document}

\markboth{C.J.~Monahan}
{Lattice Perturbation Theory and B Physics}

%%%%%%%%%%%%%%%%%%%%% Publisher's Area please ignore %%%%%%%%%%%%%%
\catchline{}{}{}{}{}
%%%%%%%%%%%%%%%%%%%%%%%%%%%%%%%%%%%%%%%%%%%%%%%%%%%%%%%%%%%%%%%%%%%

\title{THE BEAUTY OF LATTICE PERTURBATION THEORY: THE ROLE OF LATTICE
PERTURBATION THEORY IN B PHYSICS}

\author{\footnotesize C.J.~MONAHAN}

\address{Department of Physics, College of William and Mary, \\
Williamsburg, VA 23187, USA\\
cjmonahan@wm.edu}

\maketitle

%\pub{Received (Day Month Year)}{Revised (Day Month Year)}

\begin{abstract}
As new experimental data arrive from the LHC the
prospect of indirectly detecting new physics through precision tests of the Standard Model
 grows more exciting. Precise experimental and theoretical inputs are
required to test the unitarity of the CKM matrix and to search for new physics effects in rare 
decays. Lattice QCD calculations of nonperturbative inputs have reached a precision at the level of a few percent; in many
cases aided by the use of lattice perturbation theory. This review
examines the role of lattice perturbation theory in $B$ physics calculations on the lattice in the context of two questions: how is lattice perturbation theory used in the different heavy quark
formalisms implemented by the major lattice collaborations? And what role does lattice
perturbation theory play in determinations of nonperturbative contributions to the physical processes at
the heart of the search for new physics? Framing and addressing these questions reveals that
lattice
perturbation theory is a tool with a spectrum of applications in lattice $B$ physics. %% 172 words!!!
\end{abstract}

\ccode{PACS Nos.: 11.15.Ha,12.38.Bx,12.38.Gc}

\section{Introduction}

Clarifying our understanding of $B$ physics is an increasingly important undertaking in the hunt for
Beyond the Standard Model (BSM) physics.
The most recent round of
experimental results from BaBar and the LHC have imposed strict bounds on
possible BSM contributions to rare $B$ decays
\cite{albrecht12,buras12,lhcb12a,atlas12a,cms12a,becirevic12,lhcb12b,babar12a}, whilst continued
experimental and
theoretical progress has lead to more exacting tests of
Cabibbo-Kobayashi-Maskawa (CKM) matrix unitarity
\cite{lunghi11a,laiho12,pdg12,ckmfitter12,utfit12}.

Flavor-changing neutral current (FCNC) processes are forbidden at
tree-level in the Standard Model of particle physics. Such processes can
only proceed through loop contributions, which are sensitive to much higher
energy scales than the $b$ quark mass. FCNC decays therefore serve as a probe for BSM physics
and can tightly constrain the
nature and size of BSM interactions. In particular, precise
measurements of the branching fractions of the rare $B_{(s)}\rightarrow \mu^+\mu^-$
decays offer two of the most promising
avenues for the detection of BSM effects. These
decays are strongly sensitive to the existence of BSM
particles in a variety of BSM scenarios
(see, for example, the discussion and references in
\cite{albrecht12}). Semileptonic FCNC processes, such as
$B\rightarrow K^{(\ast)}\mu^+\mu^-$, may provide bounds on classes of BSM physics that are not
ruled
out by purely leptonic modes
\cite{becirevic12,altmannshofer12,altmannshofer11,bobeth10}.

Taking a different tack, the $B$ sector also provides the opportunity for detecting BSM physics
through precision
tests of the unitarity of the CKM matrix. This matrix, which is unitary in the Standard Model,
relates the mass and
weak-interaction eigenstates of down-type
quarks and incorporates all the flavor-changing and CP-violating couplings
in the Standard Model. A wide array of channels probe the
CKM matrix, such as the semileptonic decays $B\rightarrow
\pi\ell\nu$ and $B\rightarrow D^{*}\ell\nu$, and neutral $B$ mixing.
Combining independent determinations of each CKM parameter over-constrains
the elements of the CKM matrix and discrepancies or deviations from unitarity may indicate the
breakdown of the CKM framework.
Currently a number of tensions at the $2-3\sigma$ level
in unitarity fits hint at the exciting possibility of BSM physics
\cite{lunghi11a,laiho12}.

Whether the hunt for BSM physics involves FCNC processes or CKM unitarity
tests, one thing
is clear: lattice quantum chromodynamics (QCD) is an indispensable tool in
the search. Precision tests of the Standard Model require both precise
experimental and theoretical inputs. Quarks are confined to color-singlet
states and the physics of the weak interactions underlying FCNC decays and
CKM-probing processes must be teased out from the dominant nonperturbative
dynamics of the strong force. Lattice QCD is one of the most important tools presently
available for
precisely determining nonperturbative effects, but until recently has, in
many cases, been playing catch-up with experimental precision. Over the
last five years, however, the era of precision lattice QCD has dawned, with
many results now at the few percent level
\cite{daldrop11,dowdall12a,na12,bailey12a,bailey12b,bazavov12,dimopoulos12,blossier10a,blossier11}.
With the convergence of experimental and theoretical precision, and with growing
experimental datasets and advancing lattice computations, tests of the
Standard Model have become ever more exacting.

What is less clear, perhaps, is how lattice perturbation theory
fits into the march toward high precision tests of the Standard Model.
What is lattice perturbation theory? And what exactly is the
role of lattice perturbation theory in the
hunt for BSM physics? This review will answer these questions.

Viewed from the outside, Lattice QCD calculations are often seen as black boxes, simply
nonperturbative number-crunchers that spit out \emph{ab initio} predictions. Closer inspection
reveals, not surprisingly, that this is not the whole story. I will attempt to lift the lid on
heavy quark lattice QCD calculations --- at least just a little --- by examining the role of
lattice perturbation
theory in lattice B physics. I will frame this discussion in two ways: the first is the use
of lattice perturbation theory in different formulations of heavy quarks on the lattice. Viewed
through this lens, we see that lattice perturbation theory is a tool that is used, to varying
degrees, by all the major lattice collaborations working on $B$ physics. The second is the
role of lattice perturbation theory in the calculation of specific physical processes,
particularly key applications in the search for BSM physics. I focus on
precision $B$ physics; to this end I discuss not only processes
currently in the headlights of
BSM physics hunters, but also precision $B$ spectroscopy, which has proved to be a trustworthy
testing ground for lattice QCD and provides firm evidence that lattice QCD is indeed becoming more
precise and reliable. For brevity I concentrate on the impact of lattice perturbation theory in lattice $B$ physics,
rather than discussing specific details of individual calculations.

I start by motivating lattice perturbation theory in
Sec.~\ref{sec:lpt} and then briefly review recent methods for automating lattice perturbation
theory in Sec.~\ref{sec:autlpt}. In Sec.~\ref{sec:lptlimit} I examine the ways in which lattice
perturbation
theory is used in various heavy quark formulations on the lattice. I
briefly discuss relativistic formulations for heavy quarks on the lattice, but focus on
the historically more prevalent effective field theory approach and
use this discussion to limn the limits of 
lattice perturbation theory. In the second half of the review I highlight some
applications of lattice
perturbation theory to specific physical processes at the forefront of BSM searches and conclude
with a short summary in
Sec.~\ref{sec:conc}.

\section{Lattice perturbation theory}\label{sec:lpt}

Lattice QCD is the preeminent approach for \emph{ab initio} calculations
of QCD processes and is conspicuously a tool for nonperturbative physics
computations. In this view, the phrase ``lattice perturbation theory'' has the air of a
contradiction --- the perturbation theory of nonperturbative physics. But the phrase is
slightly misleading: lattice
perturbation theory is better viewed as ``perturbation theory for lattice
actions''. This move is not just cosmetic; the latter terminology is
not only less outwardly troubling, but also generally more accurate and
specific in scope. Unfortunately the new nomenclature is cumbersome;
throughout this review I will continue to simply refer to lattice
perturbation theory (henceforth ``LPT''), with the understanding that what is really at
the heart of the
discussion is ``perturbation theory for lattice
actions''.

\subsection{Motivating LPT}

A glance at the LPT literature reveals a range of applications within lattice QCD:
calculating the renormalization parameters of bare lattice actions;
matching regularization schemes and extracting continuum
results from lattice data; and improving lattice actions. These
applications, however, are inter-related and can be broadly categorized as
accounting for the physics of energy scales excluded by the lattice
cutoff. In other words, each of these applications falls under the banner
of renormalization. 

Understanding these motivations for using LPT --- 
that is, as a tool for renormalization --- is also the
key to understanding the justification for using lattice perturbation
theory, laid out in \cite{lepage96}. The lattice serves as an
ultraviolet regulator discretizing spacetime and excluding all momenta greater than $\pi/a$ (where
$a$ is the lattice spacing). As we would in any
other regularization scheme, we must calculate the renormalization
parameters of the regularized theory to correctly account for high energy effects. In
this case the excluded scales are those
above the lattice cutoff, which corresponds to approximately 5 GeV for current
lattice spacings. At these energy
scales, the coupling constant is relatively small, $\alpha_s(\pi/a)\sim
0.2$, and perturbative approximations (and therefore LPT) are likely to be
valid. 

LPT thus provides the connection
between the low and high energy regimes of QCD. Moreover, this connection has been tested and
validated in a wide range of QCD processes by comparing higher order perturbative calculations
with nonperturbative computations in the weak coupling regime
\cite{lepage00,direnzo01,horsley02,trottier02,hart04,wong06,allison08}.

There is, however, some subtlety to this story: we must choose our
coupling constant carefully to ensure the
perturbative series is well-behaved. Early LPT
calculations were plagued by slow convergence and inconsistent results
\cite{lepage93}. These issues were the product of a poor choice of
expansion parameter: the bare lattice coupling. We can greatly improve the convergence of our
perturbative series by introducing an improved
coupling constant, often defined in the ``V-scheme'' \cite{brodsky83},
expressed at an appropriately chosen scale, the ``BLM scale'' \cite{brodsky83,hornbostel02}. In
some cases,
LPT is simply insufficient and nonperturbative renormalization is required (see Section
\ref{sec:hqet} for an example).

Before turning to look at some of the uses of LPT in different lattice
formulations, I will examine in a little more detail both renormalization and improvement in the
context of LPT.

\paragraph{Renormalization and matching calculations} In general one has two choices for calculating
renormalization parameters, matching parameters or improvement 
coefficients for lattice actions: nonperturbative tuning or perturbative calculation. There are a
variety of nonperturbative tuning methods. A full review is beyond the scope of this report, but a
partial list includes step-scaling methods, which may be 
applied to the Schr\"odinger functional \cite{luescher91}, off-shell Green functions (the
``Rome-Southampton method'')\cite{martinelli95} or physical quantities\cite{lin07}; imposing Ward
identities or chiral symmetry relations \cite{luescher96}; and fixing to physical quantities, such
as meson masses, or relations, such as relativistic dispersion relations \cite{aoki12}.

Although these approaches differ considerably in practice, they share two particular advantages. All
nonperturbative methods enable the replacement of
perturbative truncation errors, which can be hard to quantify, with statistical and systematic
errors, which can usually be determined reliably. In general, with intensified computational
effort, these uncertainties can be systematically reduced. Secondly, a fully nonperturbative
approach is the only truly \emph{ab initio} method for
QCD calculations.

Nonperturbative methods share a number of disadvantages as well: 
step-scaling and iterative methods incur a greatly increased computational expense, whilst matching
to physical quantities results in a loss of predictive power. 

LPT is an alternative to nonperturbative tuning.
In this approach one calculates the renormalization constants or matching parameters
perturbatively, carrying out the calculation in a spirit broadly similar to
the standard perturbative calculations of continuum QCD. LPT is computationally much cheaper than
nonperturbative
tuning and there is no loss of predictive power. There are disadvantages, too, of course. Most
notably LPT introduces perturbative truncation errors. As lattice calculations become
more precise, the
concomitant perturbative error often become the largest source of
uncertainty in the result, which will ultimately necessitate multi-loop
calculations will eventually be necessary. This is no mean feat: LPT is usually more
involved than the corresponding calculations in continuum QCD. The
relatively recent advent of
sophisticated automated perturbation routines for lattice actions, however,
means that multi-loop calculations for improved actions have
been
carried out \cite{nobes02,mason05,mason05b,hart10}. Despite the complications, the computational
cost of such calculations is invariably lower than that for nonperturbative
tuning, ensuring LPT is an attractive alternative to nonperturbative
renormalization.

\paragraph{Improvement} Precise lattice
QCD computations generally require lattices with two properties: small
lattice spacings to reduce discretization errors and large lattice volumes
to remove finite size effects. Unfortunately lattice QCD calculations tend to be
rather computationally expensive. And as the lattice spacing decreases and
the lattice size increases, they grow ever more expensive. 

Thankfully for
precision $B$ physics there is an alternative to simply using ever finer lattice spacings. Lattice
actions can instead be ``improved'' to eliminate
sources of error. The most common approach to improvement is the Symanzik
improvement programme \cite{symanzik83a,symanzik83b}, in which scaling
violations are systematically removed by adding irrelevant operators to the
lattice action, organized by mass dimension. This approach is frequently significantly cheaper than
simulating at smaller lattice spacings.

Determining the coefficients of the new terms in the improved action can usually
be carried out quite simply at tree-level, but for precision studies, this
is insufficient. By construction lattice actions exhibit the same long-range physics
as continuum QCD, but their short-range behaviour is distorted by the
lattice cutoff. Radiative corrections to the new operators in the action
renormalize the coefficients away from their tree-level values and this
can lead to new sources of uncertainty in lattice
predictions. By matching onshell quantities on the lattice and in the continuum, the resultant
discrepancies can be removed order-by-order in perturbation theory.

There is naturally a price to be paid for adding extra terms to the action (radiatively improved or
otherwise). The extra operators considerably
complicate the Feynman rules for the improved actions. Highly improved
actions, such as the HISQ and NRQCD actions that I discuss in Sec.~\ref{sec:lptlimit}, generate
Feynman
rules that cannot
feasibly be manipulated by hand. Furthermore, the lack of Lorentz
symmetry complicates the evaluation of Feynman integrals, which are
no longer amenable to Feynman parameter transformations or the other
tricks of the continuum trade and must be handled numerically. To
cope with these complications LPT has been
automated.

\section{LPT today: Automation}\label{sec:autlpt}

In recent years a number of automated lattice
perturbation theory algorithms have been developed, many following the pioneering work of L\"uscher
and
Weisz\cite{luescher86}. An early variant of the L\"uscher-Weisz (LW) algorithm 
was deployed in \cite{nobes02,nobes03,nobes04}, but currently the
most widely used descendant of the LW algorithm 
is implemented in the \hippy$\!\!$/\hpsrc software packages \cite{hart05,hart09}. More recently,
the LW algorithm was adapted for gauge actions in the Schr\"odinger functional scheme
\cite{takeda09}.
A new software package, called \pastor, has extended this work to include Wilson-type
relativistic quarks and HQET heavy quarks \cite{hesse11}. Initial calculations
with \pastor are underway
\cite{hesse12a,hesse12b}.
Finally,
following an altogether different approach, a computer algebra system has been
constructed and optimized for LPT
\cite{lehner12}. Although the use of this software has so far been
restricted to a calculation using the relativistic heavy quark action in
the Columbia formulation \cite{aoki12}, in principle the algorithm
can be extended to any lattice quark action.

The \hippy$\!\!$/\hpsrc software packages have now been used in a variety of
perturbative calculations, for example in
\cite{dowdall12a,hart04,hart07,mueller11,drummond02,drummond03,hammant11}, and extensively tested against previous results.
 Evaluating Feynman integrals is a two stage process with
the \hippy$\!\!$/\hpsrc routines: one first generates the Feynman rules
with \hippycomma a set of \python routines that encode the Feynman rules in ``vertex files''. These vertex files are
then read by the
\hpsrc \fortran modules, which evaluate the corresponding integrals numerically via
the \vegas algorithm \cite{lepage80} or exact mode summation over a finite lattice. Any
derivatives required in the
calculation are implemented analytically using 
the derived \verb+taylor+ type, defined as part of the \fortran
\taylor package \cite{hippel10}. Both \hippy and \hpsrc implement parallel processing using MPI
 (Message Passing Interface). 

The Schr\"odinger functional scheme has been widely used over the last thirty years (see the
Sec.~\ref{sec:hqet} for references). The advent of
LPT for the Schr\"odinger functional scheme is, however, much more recent. There are two
reasons for this: the first is the additional
complexity introduced by the Schr\"odinger functional formalism and the
second is that LPT has very limited
application to HQET computations, as I discuss in more detail in the next
section. LPT is nevertheless useful, because
perturbative calculations can serve as a starting point for more precise
nonperturbative methods. These nonperturbative methods are
computationally expensive and there are a large number of renormalization
parameters in the Schr\"odinger functional scheme, many of which have only
a small relative effect on the final result. Perturbative investigations
can guide where computational resources should be
focussed to most improve the precision of lattice calculations.

There are three complications that must be handled by automated routines
for the Schr\"odinger functional scheme. Firstly, translational invariance,
which is assumed in the \hippy$\!\!$/\hpsrc algorithm, is broken in the
time direction. Secondly, we must account for an induced abelian color background
gauge field\footnote{In fact, the \hippy$\!\!$/\hpsrc
packages have been extended to use background field gauge 
\cite{hammant11}, but this implementation is not sufficient for the 
Schr\"odinger functional scheme.}. Thirdly, at a given order in the
coupling constant, there are extra diagrams generated by the boundary
conditions of the Schr\"odinger functional scheme that are not present in
other lattice formulations. The \pastor software package, written specifically to deal with these
difficulties, is based on
\texttt{C++} routines that generate vertices and a \python
frontend wrapper that allows the user to specify the lattice actions in
symbolic form in \texttt{C++}. An \texttt{xml} input file specifies the desired observable in
the Schr\"odinger functional scheme, which the \python frontend routine parses
to generate all the relevant diagrams of order $g_0^2$. These diagrams are then evaluated with a
\texttt{C++} program generated by the \python wrapper \cite{hesse11,hesse12a}.

The most recent automated LPT framework, presented in \cite{lehner12}, is
based on a new \texttt{C++} computer algebra library optimized for lattice
perturbation theory. Lattice actions are defined by \texttt{C++} classes that allow the user to
specify the action in text form. These classes extract the vertices, then further \texttt{C++}
classes perform Wick contractions and convert the algebraic representation of the integrand into
efficient \texttt{C++} code that can be evaluated numerically. This framework also contains routines
that can perform analytic differentiation with respect to the external momenta and
undertake continuum calculations in dimensional regularization using the Passarino-Veltman
reduction \cite{passarino79}.

\section{LPT today: Applications in heavy quark lattice
formulations}\label{sec:lptlimit}

The computational expense of lattice QCD has been a distinct problem for $b$ quarks, which
are heavy and therefore ``fall through'' lattices
that are too coarse to accurately resolve the Compton wavelength of the $b$ quark.
Putting this more precisely: it is not currently computationally feasible
to use lattices large
enough and fine enough to simultaneously satisfy $m_\pi L \gg 1$ and $m_ba
\ll 1$. Here $m_\pi$ is the mass of the lightest propagating particle in
the theory; $L$ is the side length of the lattice; and $m_b$ is the mass of
the $b$ quark. Heavy-light systems, which feature heavily in BSM searches in the $B$ sector,
represent the worst of all computational worlds; they are
characterized by both the mass of the light quark and the $b$ quark
--- scales separated by several orders of magnitude. 

Lattice QCD has traditionally dealt with this problem using the technology of effective field
theories. Despite the considerable advances in computing infrastructure
and simulation algorithms over the
last thirty years, even state-of-the-art lattice computations require
extrapolations up to the $b$ quark mass \cite{dimopoulos12,mcneile10,mcneile12a,mcneile12b}
and effective field theories continue to play an
important part in lattice simulations.

Historically, the application of effective field theoretic techniques to
lattice QCD has elucidated the interplay of
perturbative and nonperturbative physics and has consequently helped
advance the understanding of conceptual issues such as the
operator product expansion and the role of
renormalons. I will examine effective theories for heavy quarks more closely
to highlight some of the limitations of LPT,

There are two widely used approaches to heavy quarks on the lattice:
nonrelativistic QCD (NRQCD) and heavy quark effective theory (HQET). Both
take advantage of the large mass of the $b$ quark relative to other scales
in QCD physics and both have the same infinite mass, or stationary quark,
limit. There are, however, important differences. Here I only briefly outline some of the important
considerations for LPT for both
lattice NRQCD and HQET. For a more complete pedagogical
introduction to continuum NRQCD, see, for example,
\cite{braaten97}. Reviews of HQET in the continuum appear in
\cite{neubert94} and in the textbooks \cite{manohar00,grozin04}, whilst lattice HQET is reviewed
in \cite{sommer10}.

\subsection{NRQCD}

Heavy quark bound states are typified by the small relative
velocity of their constituent quarks. In heavy $b\overline{b}$ states,
such as the $\eta_b$ and $\Upsilon$ mesons, this relative velocity is
approximately $v^2\simeq 0.1$. Consequently, $v$ induces
three well-separated energy scales: the mass, ${\cal O}(m_b)$; the
momentum, ${\cal O}(m_bv)$; and the kinetic energy ${\cal O}(m_bv^2)$. NRQCD is also suitable for
heavy-light systems, such as the $B$ and $B_s$ mesons, in which case the expansion parameter is
$\Lambda_{\text{QCD}}/m_b$ and the power counting is that of HQET, but for simplicity I restrict my
discussion to heavy-heavy systems.

To construct the NRQCD action we use the
Foldy-Wouthuysen-Tani transformation \cite{itzykson80} to decouple the quark and antiquark fields
and undertake a nonrelativistic expansion. 
The result is an effective theory with nonrelativistic, low-energy degrees of freedom. At
lowest order, the continuum NRQCD action is just the
nonrelativistic Schr\"odinger action.

Following the Symanzik improvement scheme, we can then systematically
include interactions and improve the NRQCD action to a given order in $v$
and $\Lambda_{\text{QCD}}/m_b$ by adding irrelevant higher order
operators. Finally, armed with an appropriately improved continuum NRQCD
action, we then
discretize it for lattice computations.
 We must choose the coefficients of the higher order
operators to ensure the effective lattice theory replicates
the correct behaviour for physical observables. As I discussed
in the previous section, calculating the radiative corrections to the
improvement coefficients is the role of LPT.

One property of lattice NRQCD is a particular challenge: lattice NRQCD is a nonrenormalizable
effective theory with no continuum limit. We cannot take the limit of
vanishing bare lattice mass, $am_b\rightarrow 0$. Even at leading order in
the Foldy-Wouthuysen-Tani transformation, the $1/(am_b)$ corrections are
present in the heavy quark propagator. Interaction and improvement terms
in the NRQCD action are higher order in $1/(am_b)$ and consequently
there are an infinite number of divergences that cannot be absorbed into
the renormalization parameters.

From a perturbative standpoint, this means the series that
define the radiative corrections to the higher order operators in the
action become ill-defined at small masses. LPT --- and indeed NRQCD ---
fails in the region $am_b \leq 1$. This has been explicitly demonstrated for the case of
the chromomagnetic correction to NRQCD by studying the mass dependence for a range of masses
around $am_b=1$ \cite{hammant12}. We can safely use LPT to
improve lattice NRQCD, provided we remain in the region for
which $am_b> 1$. This ensures that nonperturbative computations using
lattice NRQCD converge with those obtained from continuum QCD.

Although we can not take the continuum limit of lattice NRQCD --- it does
not exist --- we can nevertheless extract continuum results from lattice
NRQCD calculations. By carrying out computations at different values of
the lattice spacing within the region of validity of NRQCD, \emph{i.e.}~at different values of
$am_b$ for $1/m_b < a < 1/\Lambda_{\text{QCD}}$, we may
parameterize the lattice spacing dependence of our results and quote
uncertainties appropriately. With sufficient improvement, we can reduce
the lattice spacing dependence to, for example ${\cal O}(1/(am_b)^2)$ or
higher, and obtain correspondingly precise results, often at the few
percent level. Without LPT, such precise
nonperturbative results for many $B$ physics parameters would
be a distant hope for the future.

\subsection{HQET}\label{sec:hqet}

Although HQET bears many similarities to NRQCD, the approach is both
conceptually and practically distinct. The first conceptual difference is that, unlike
NRQCD, HQET is an effective theory for singularly heavy hadrons; mesons such as the $\Upsilon$
or $\eta_b$ are off limits. The second crucial distinction is that HQET is believed to be renormalizable
at leading order, \emph{i.e.}~in the infinite heavy quark mass, or
``static'', limit. The third contrast is that, even in the static limit, the
renormalization and matching parameters of lattice HQET must be calculated nonperturbatively
for precise $B$ physics results \cite{heitger04}. LPT will not do.

HQET is motivated by the intuition that singularly heavy hadrons are analogous to a
hydrogen atom, with the $b$ quark playing the part of the proton and the light quark the electron.
Constructing HQET
follows the same effective field theory approach to building NRQCD: we identify the relevant
degrees of freedom, decouple the high and low energy modes via the Fold-Wouthuysen-Tani
transformation and
include all terms consistent with the desired symmetries at a given order in the expansion. 
We start by building an effective field theory that describes the light
degrees of freedom interacting with a static color source, the heavy quark. Then, rather than
including higher order operators directly in the HQET Lagrangian, we expand the
non-leading terms in the path integral weight factor, $\exp(-S_{\text{HQET}})$, so that the
non-leading contributions appear as insertions in correlation functions. More explicitly, we
write 
\begin{align}
\exp(-S_{\text{HQET}}) = {} & \exp\bigg[-a^4\sum_x\left({\cal
L}^{\text{static}}_{\text{HQET}}(x)+
{\cal L}^{(1)}_{\text{HQET}}(x)\right)\bigg] \nonumber \\
= {} & \exp\bigg[-a^4\sum_x{\cal
L}^{\text{static}}_{\text{HQET}}(x)\bigg]\left(1-a^4\sum_x{\cal
L}^{(1)}_{\text{HQET}}(x)\right),
\end{align}
where ${\cal L}^{\text{static}}_{\text{HQET}}$ is the leading order HQET Lagrangian and ${\cal
L}^{(1)}_{\text{HQET}}$ contains higher order terms. This approach ensures that HQET remains
renormalizable and the continuum limit of HQET correlation functions exists, provided all local
operators at a given order in $1/(am_b)$ are included.

The parameters of the effective theory are fixed at tree-level by the FWT transformation, but
beyond tree-level these must be found by matching to a $1/m_b$ expansion of continuum QCD.
Perturbative matching is insufficient for precision $B$ physics. Continuum HQET
matching factors have been perturbatively studied in, for example, \cite{bekavac10},
where it was demonstrated that even at three-loops the convergence of the perturbative series
is very slow. The matching coefficients are scale dependent, and their magnitude can be
reduced by lowering the scale, but this does not help: at the scale at which the perturbative
series converges quickly the coupling constant itself is rather large. There
seems to be no escaping the conclusion that for reliable results at the precision of a few
percent, nonperturbative matching is necessary \cite{sommer10}.

Beyond leading order, matrix elements of higher dimensional operators mix and their
coefficients must be fine tuned. These operators diverge with inverse powers of the lattice
spacing and attempting to remove these power divergences via perturbative matching introduces
ambiguities associated with so-called ``renormalons''. A full study of renormalons and their
role in HQET would take us too far from our central topic; for in-depth reviews with applications to $B$ physics and
HQET, see for example, \cite{sachrajda96,beneke99,grozin03}. In essence renormalons arise because
perturbative QCD is an asymptotic approximation to QCD that fails to capture
nonperturbative behavior. A common example is the pole mass, which is a long distance quantity
that can be rigorously defined order-by-order in perturbation theory as the pole of the
renormalized quark propagator, but is subject to inescapable renormalon ambiguities of ${\cal
O}(\Lambda_{\text{QCD}}/m_b)$ when nonperturbative effects are included \cite{bigi94,beneke94}.

For our purposes, though, the central point is that renormalization parameters of HQET must be
calculated nonperturbatively. This is a non-trivial task to implement for precision
$B$ physics, because one must ensure firstly that the HQET expansion is sufficiently accurate
and secondly that the numerical precision is adequate. On top of this, the lattice simulations
must be undertaken with physical volumes large enough to avoid significant finite volume
corrections. Currently the ALPHA collaboration is carrying out a programme of precision
$B$ physics using HQET for the $b$ quark
\cite{blossier10a,blossier11,dellamorte07,blossier10b,blossier10c} and implementing
nonperturbative renormalization using the Schr\"odinger functional scheme and a step-scaling
method \cite{blossier10d,blossier12}.

All this does not mean, however, that LPT has no role to play in HQET
computations. In fact, the ALPHA collaboration has recently developed automated LPT routines for the
Schr\"odinger functional scheme \cite{hesse11,hesse12a} to serve as
an exploratory guide to focus computational effort most efficiently in nonperturbative
determinations.

\subsection{Relativistic heavy quark actions}

Relativistic heavy quark actions combine aspects of effective theories and fully
relativistic treatments and hence serve as a suitable framework for both light and heavy quarks
\cite{elkhadra97,aoki03,christ07a}. The basic idea of the relativistic heavy quark formalism is
to extend the Symanzik effective theory approach, on which NRQCD and HQET are based, to
include interactions from both the small $m_qa$ and large $m_q/\Lambda_{\text{QCD}}$ limits. As
we take the limit of vanishing bare mass, $m_qa\rightarrow 0$, the relativistic heavy quark
action reduces to the ${\cal O}(a)$-improved clover action \cite{sheikholeslami85}. Conversely,
in the large bare mass limit, when $m_q \gg \Lambda_{\text{QCD}}$, the action breaks the
time-space axis interchange symmetry and can be interpreted nonrelativistically, with a universal
static limit.

Both
the Fermilab Lattice/MILC collaboration and the RBC/UKQCD collaboration, currently employ
relativistic heavy quarks for precision $B$ physics \cite{bazavov12,aoki12}. These  collaborations'
approaches to the relativistic heavy quark formalism differ primarily in the methods they use to
tune their actions.

The RBC/UKQCD method uses three physical conditions to tune their parameters, matching to the
spin-averaged $B_s$ mass and hyperfine splitting and ensuring the continuum dispersion relation
holds for the $B_s$ meson. This tuning process is nonperturbative, but there is nonetheless a role for
LPT. Perturbative calculations of the three tuning parameters
at one-loop have been carried out to provide
a consistency check for the nonperturbative tuning process \cite{aoki12,lehner12}. Ultimately LPT will be used
for parameters for which nonperturbative tuning is not available.

In the Fermilab approach, two of the parameters are fixed and the third tuned nonperturbatively,
using the spin-averaged $B_s$ mass \cite{bernard11}. LPT played its
part in the original construction of the Fermilab action \cite{elkhadra97}, and remains an
important component for matching lattice matrix elements to their continuum counterparts. For
decays such as the semileptonic $B\rightarrow \pi\ell\nu$ and $B\rightarrow D\ell\nu$ decays, the
Fermilab Lattice/MILC
collaboration use a mixed approach to matching. The bulk of the matching is carried
out nonperturbatively, leaving a factor close to unity that can be
calculated perturbatively \cite{harada02a,harada02b,harada05,elkhadra07}. For other
processes, such as neutral $B$ mixing, however, the Fermilab Lattice/MILC collaboration uses a
purely perturbative matching procedure \cite{bazavov12}.

Finally, it is worth noting that LPT played a critical role in the formulation of the
relativistic heavy
quark action used by the Tsukuba group \cite{aoki03}. In this formulation, there is an extra
parameter in the action that cannot be nonperturbatively determined \cite{aoki04}.

\subsection{Fully relativistic $b$ quarks: HISQ and twisted mass fermions}

Two lattice collaborations have
recently begun to implement a programme of precision $B$ physics using purely relativistic
actions: the HPQCD collaboration \cite{mcneile10,mcneile12a,mcneile12b} and the ETM collaboration
\cite{dimopoulos12,blossier09,blossier10e}.

These collaborations use different relativistic actions in their computations, but both
presently require extrapolations up to the physical $b$ quark mass. The HPQCD collaboration uses
the highly improved staggered quark (HISQ) action \cite{follana07}, whilst the ETM collaboration
employs a twisted mass variant of the Wilson action. The HISQ action exhibits exact chiral symmetry
in the massless limit and
therefore the heavy-light axial-vector and vector currents are
absolutely
normalized and do not require operator matching. The role of LPT in
HISQ calculations is
consequently somewhat reduced, but not completely absent. One loop calculations were required in the construction of the HISQ action. Furthremore, 
renormalization
parameters for the tensor current and four quark operators are not
absolutely normalized and therefore require matching to their continuum counterparts.

The ETM collaboration uses nonperturbative renormalization for the quark bilinear operators used
in $B$ physics applications \cite{constantou10}. Nevertheless, LPT is used to parameterize the
discretization errors associated in the scale dependence of
the nonperturbative renormalization factors in the RI-MOM scheme
\cite{constantou12}. Subtracting the
one-loop perturbative behaviour significantly improves the size of discretization errors in
the renormalization constants \cite{constantou09}. The part that LPT
plays here may be restricted, but it has still been important in improving the precision of
twisted mass computations.

As above discussion makes clear, even as nonperturbative simulations move closer to the ideal of
true \emph{ab initio} calculations of QCD processes, LPT retains a vital
role in extracting precise results. To clarify this role further, I now turn to the application
of LPT in specific $B$ physics processes.

\section{LPT today: Applications in $B$ physics}\label{sec:bapp}

A handful
of key $B$ processes have become the loci of considerable
experimental and theoretical attention. Effort has naturally focused on
those processes that have, or are likely to soon have, small experimental
and theoretical uncertainties. I will concentrate on only those
channels that lie at the heart of the searches for BSM
physics and CKM matrix unitarity violation and restrict my discussion to
exclusive channels. Inclusive processes, whilst playing a vital role in
BSM searches and contraining CKM matrix unitarity, fall outside the scope
of this short review.

I summarize the processes in which LPT has
been an important ingredient in Table \ref{tab:processes}. I break the
table into three sections: in the top third I tabulate the key processes
currently in the headlights of the flavor physics community. In the
second third I present other processes for which LPT has been necessary for precise nonperturbative
results. These
processes have yet to achieve the experimental or
theoretical importance of the processes listed in the first third of the
table (usually because of larger experimental or theoretical
uncertainties), but are nevertheless playing an increasingly important
role in flavor physics in the heavy quark sector. In the final third, I present some
more speculative approaches to uncovering new physics. Such processes are either poorly understood
theoretically or are yet to be observed experimentally.

Naturally this 
hierarchy is largely a difference of degree rather than
difference in kind, but highlights the crucial processes at the center of current BSM
searches. Within Table \ref{tab:processes} I
distinguish between processes that determine CKM matrix elements, and
thereby constrain CKM matrix unitarity, and rare decays that offer
hope for detection of BSM loop effects. I represent this distinction in the second column. In
the third column I give the quantity determined by lattice QCD that
characterizes the nonperturbative physics of the process. In column
four I list the relevant renormalization parameters calculated using LPT. I use the shorthand $Qq$ to
represent heavy-light currents, $QQ$ for heavy-heavy currents and $QQqq$ to indicate
four-fermion interactions.

\begin{table}[h]
\tbl{Summary of $B$ physics processes discussed in this article. For
a full description of the table, see the accompanying text.}
{\begin{tabular}{@{}ccccc@{}} \toprule
Process & $B$ physics & Lattice parameter &
Renormalization (LPT) & Refs. \\ 
\colrule
$B\rightarrow \pi\ell\nu$ & $V_{ub}$ & form factors & $Qq$
vector current & \cite{mueller11,elkhadra07,monahan12,gulez04}\\
$B\rightarrow D^{*}\ell\nu$ & $V_{cb}$ & form factors
& $QQ$ vector current & \cite{elkhadra07,monahan12,kuramashi98,boyle00} \\
$B_s\rightarrow\mu^+\mu^-$ & rare decay & decay constant & $Qq$ axial-vector
current & \cite{elkhadra07,monahan12,gulez04} \\
$B_q^0-\overline{B^0_q}$ mixing & $V_{td}/V_{ts}$ &
$SU(3)$ breaking ratio &
$QQqq$ operator & \cite{loktik07,gamiz08,hashimoto00} \\
\colrule
$B\rightarrow \tau \nu$ & $V_{ub}$ & decay constant & $Qq$ axial-vector
current & \cite{elkhadra07,monahan12,gulez04}\\
$B\rightarrow D\ell\nu$ & $V_{cb}$ & form factors
& $QQ$ vector current & \cite{elkhadra07,monahan12,kuramashi98,boyle00}\\
$B\rightarrow K\ell^+\ell^-$ & rare decay & form factors & $Qq$
vector current & \cite{mueller11,monahan12,gulez04}\\
$B\rightarrow K^\ast\gamma$ & rare decay & form factors & $Qq$
tensor current & \cite{mueller11}\\
\colrule
$B\rightarrow X_u\ell\nu$ & $V_{ub}$ & form factors & $Qq$
vector current & \cite{mueller11,elkhadra07,monahan12,gulez04}\\
$B_s\rightarrow K\ell\nu$ & $V_{ub}$ & form factors & $Qq$ vector current & \cite{mueller11,monahan12,gulez04} \\
$B_c\rightarrow \tau \nu$ & $V_{cb}$ & decay constant & $QQ$ axial-vector
current & \cite{elkhadra07,monahan12,kuramashi98} \\
\botrule
\end{tabular}\label{tab:processes} }
\end{table}

\subsection{Heavy-light current renormalization}\label{sec:lptcurrent}

Matrix elements of quark currents are the starting point for any lattice
computation of nonperturbative QCD contributions to weak interaction
processes. Understanding quark current renormalization is consequently a
critical component of precise lattice QCD predictions.

The vector current mediates weak interactions between states of the same
parity, such as the semileptonic decay $B\rightarrow
\pi \ell \nu$, which is parameterized by
\begin{equation}
\left\langle \pi | V^\mu |B\right\rangle =
f_+(q^2)\left[p_B^\mu+p_\pi^\mu-\frac{m_B^2-m_\pi^2}{q^2}q^\mu\right]
+f_0(q^2)\frac{m_B^2-m_\pi^2}{q^2}q^\mu,
\end{equation}
where $f_+$ and $f_0$ are the ``form factors''. In the limit
$m_\ell\rightarrow0$, a good approximation for electron and muon
neutrinos, the decay rate reduces to
\begin{equation}
\frac{\mathrm{d}\Gamma}{\mathrm{d}q^2} \propto |V_{ub}|^2|f_+(q^2)|^2.
\end{equation}
This channel lies at the center of one of
the main tensions in analyses of CKM matrix unitarity: currently, the
values of $V_{ub}$ from inclusive and exclusive semileptonic decays exhibit
a $3.3\sigma$ disagreement \cite{laiho12}. 

Other semileptonic $B$ modes, such as $B\rightarrow \rho \ell\nu$, $B\rightarrow \eta\ell\nu$ and
$B\rightarrow \omega\ell \nu$, represent rather more speculative possibilities for the extraction
of $V_{ub}$. The study of such decays is currently in its infancy for lattice QCD (see, for
example, \cite{wingate06} for a discussion of some of the challenges), but in principle any
independent measurement offers the opportunity for greater understanding of the CKM mechanism.
 Similarly, the $B_s\rightarrow K\ell\nu$ provides another exclusive determination of $V_{ub}$ and is
 currently being explored on the lattice \cite{bouchard12},
in advance of experiments proposed at LHCb \cite{bozzi12} and Belle \cite{urquijo12}.

Purely leptonic decays of pseudoscalars, such as $B\rightarrow\ell\nu$,
are directly sensitive to $V_{ub}$ and therefore offer further insight into
this tension. Such decays
require a spin-flip, however, and the decay rate is helicity-suppressed by $m_\ell^2/m_B^2$. Only
the $B\rightarrow\tau \nu$ mode has been observed \cite{babar12b}. Consequently the experimental
uncertainty of the branching
ratio prevents determinations of $V_{ub}$ from reaching the precision
achieved with the semileptonic modes. In addition, the leptonic decay $B\rightarrow\tau \nu$ is
expected to be sensitive to the presence of
a charged Higgs boson, which is an exciting opportunity for experimentalists, but means
that determinations of $V_{ub}$ from this process may be difficult to interpret in terms of
explicit unitarity violation. Nevertheless, experimental
uncertainties will decrease and precise lattice computations will ensure that
leptonic decays can ultimately deepen our understanding of the
$V_{ub}$ discrepancy. 

Furthermore, the decay constants that parameterize
leptonic decays are crucial ingredients in indirect searches for BSM
physics. Observation of the rare $B_{(s)}\rightarrow
\mu^+\mu^-$ decays is one of the key aims of the LHC.
The high sensitivity of these processes to BSM physics ensures that they are
amongst the star contenders for ``likeliest channel for new physics
results''. The branching fractions of these rare decays are proportional to the square of the decay
constants,
$f_{B_{(s)}}$, defined via the matrix elements of the leptonic decays
$B_{(s)}\rightarrow\ell\nu$ as
\begin{equation}\label{eq:fbs}
\left\langle 0 |A_0 |B_{(s)}\right\rangle = f_{B_{(s)}}M_{B_{(s)}},
\end{equation}
where $A_0$ is the temporal component of the axial-vector current and
$M_{B_{(s)}}$ the mass of the $B_{(S)}$ meson. Although the decay channel
$B_s\rightarrow\ell\nu$ is not a physical process, it is nonetheless a
process that can be computed within lattice QCD. In fact, the most precise
values for the branching fraction of $B_{(s)}\rightarrow
\mu^+\mu^-$ are obtained by taking a ratio with the
$B_s^0-\overline{B_s^0}$ mixing mass difference, $\Delta M_s$
\cite{buras03}. In this case, the dependence on $f_{B_s}$ drops out. Determinations using both
methods have been remarkably consistent, providing a strong check of the reliability of
lattice computations, because the lattice calculations for $B_s$ decays and $B_s-\overline{B}_s$
mixing are rather different.

\begin{figure}[h]\label{fig:fspec}
\centerline{\psfig{file=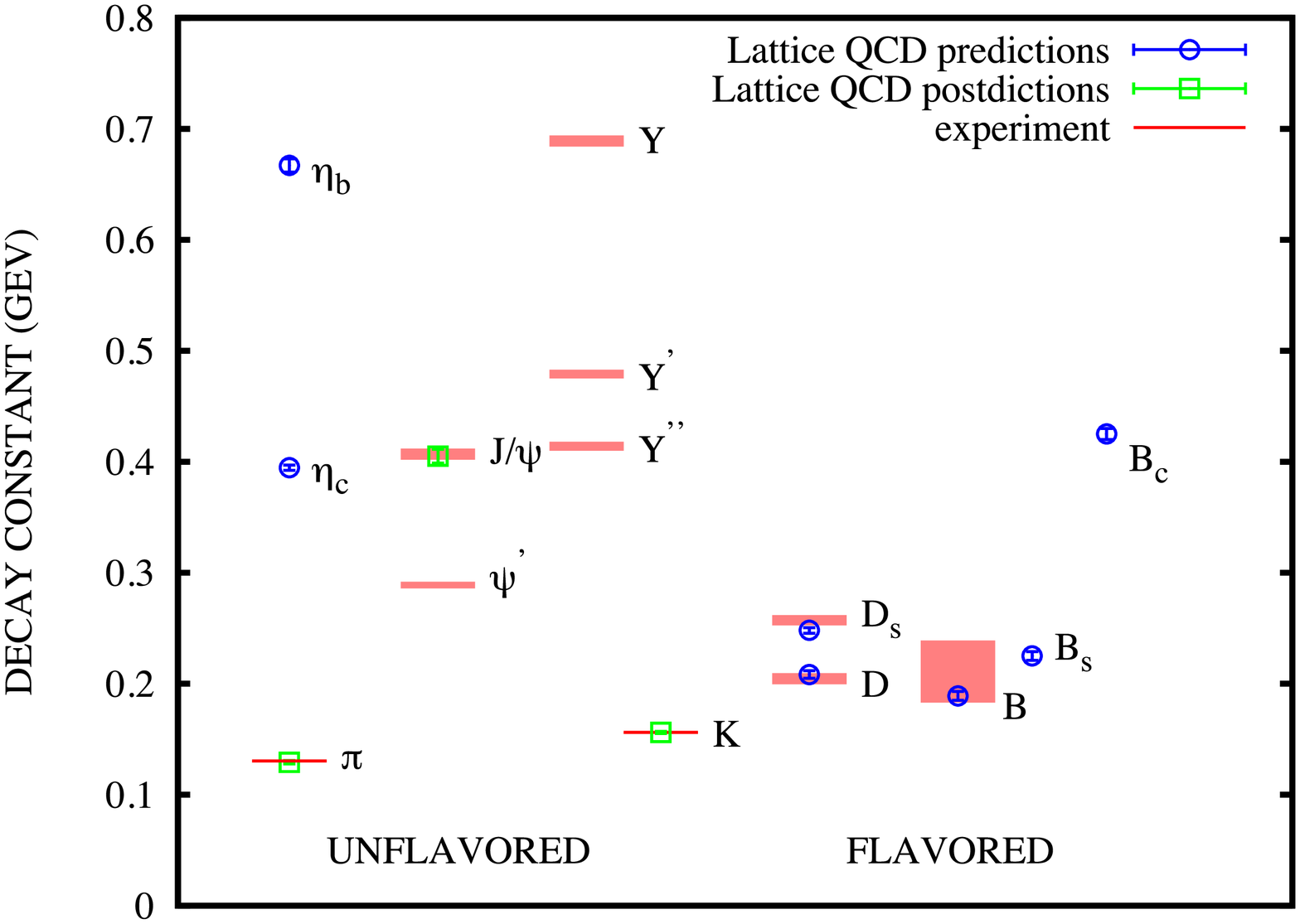,width=0.8\textwidth}}
\vspace*{8pt}
\caption{The current status of decay constants calculated by the HPQCD
collaboration. Figure reproduced courtesy of C.T.H.~Davies and the HPQCD collaboration, updated
from C. McNeile \emph{et al.}~$^{44}$.}
\end{figure}

Precise determinations of form factors and decay constants from lattice QCD
are therefore of paramount importance in the hunt for BSM physics.
And this is where LPT steps into the fray.
The most up-to-date determinations of form factors and decay constants come from four lattice
collaborations
using four
different formalisms. The ETM collaboration employ
twisted mass fermions
with $n_f=2$ species of sea quarks (with $n_f=2+1+1$ calculations
underway) \cite{dimopoulos12}, whilst early results from the ALPHA collaboration on $n_f=2$
ensembles with ${\cal O}(a)$-improved Wilson fermions are now available \cite{bernardoni12,bahr12}. The Fermilab
Lattice/MILC group \cite{bailey09,bazavov11} and
the HPQCD collaboration both use staggered fermion ensembles generated by
the MILC collaboration, with different valence quark actions. The HPQCD collaboration has
examined a wide range of flavored and unflavored meson decay constants. The current status of
the decay constant ``spectrum'' from the HPQCD collaboration is illustrated in Fig.~\ref{fig:fspec}
and demonstrates the excellent agreement with experimental data.

All four collaborations use LPT, albeit to varying degrees. The HPQCD
collaboration has made heavy use of LPT in recent computations, so I
turn now to an example for the quark currents used by the HPQCD collaboration.

\subsubsection{A physics example: LPT in the HPQCD approach}

The HPQCD collaboration has determined both the decay constants $f_B$ and
$f_{B_s}$ \cite{mcneile12a,na12,monahan12} and the form factors $f_+$ and $f_0$
\cite{gulez06,gulez07}.
These determinations fall into two classes: the more traditional
calculations use nonrelativistic $b$ quarks paired with relativistic
staggered light quarks whilst recent computations have pioneered a
completely relativistic treatment of the $b$ quarks using the HISQ action.
This entirely relativistic approach has several advantages that
significantly improve the precision of the computation; the
resulting value, $f_{B_S}^{\mathrm{HISQ}} = 0.225(4)$ GeV \cite{mcneile12a}, is the most precise
currently available. Using relativistic $b$ quarks avoids any need for
effective actions and their associated systematic uncertainties, and the HISQ action
has an absolutely normalized axial-vector current, removing the
requirement for renormalization parameter calculations.

This approach does not make LPT redundant,
however. Simulations with relativistic $b$ quarks are yet to reach the
physical point and a determination of $f_B$ is even more
computationally demanding. Lattice calculations with NRQCD remain the
most competitive method for determining the ratio $f_B/f_{B_s}$.
Systematic uncertainties associated with the
perturbative matching or with neglecting higher order NRQCD terms are
correlated between $B$ and $B_s$ simulations and therefore cancel in the
ratio. The most precise result available for $f_B$ takes advantage of
these cancellations by combining a precise determination of the ratio
$f_B/f_{B_s}$ with the precise determination of $f_{B_s}$ using HISQ $b$
quarks to obtain $f_B=0.189(4)$ GeV \cite{na12}. This result would have not been possible without
LPT.

\subsection{Heavy-heavy current renormalization}\label{sec:bheavy}

Heavy-heavy currents are required as part of the extraction of the CKM matrix element $V_{cb}$ via
exclusive semileptonic decays. The decay $B\rightarrow D^\ast \ell \nu$ is currently the most
important exclusive process for attempts to reconcile the discrepancies between determinations of
$V_{cb}$ from inclusive and exclusive channels, but $V_{cb}$ can also be extracted from
$B\rightarrow D \ell \nu$. 
The
BaBar collaboration
recently reported an excess of $3.4\sigma$ above the expected Standard Model ratio
$\text{BR}(B\rightarrow D^{(\ast)}\tau\nu)/\text{BR}(B\rightarrow D^{(\ast)}\ell\nu)$ in both
channels \cite{babar12c}, although a very recent lattice determination has reduced the tension to
less than $2\sigma$ \cite{bailey12a}.

%Heavy-heavy mesons are also of theoretical interest. In particular the $B_c^+$ meson lies below the
%$BD$ threshold and can only decay weakly; it therefore has a relatively long
%lifetime. First observed, with a poor mass resolution of only 400 MeV, by the CDF
%collaboration at the Tevatron collider in 1998 \cite{cdf98}, a later ``prediction'' of the 
%$B_c$ was a significant victory for lattice QCD. I discuss this in more detail in Section
%\ref{sec:bspec}. Furthermore. direct
Direct access to the CKM matrix element $V_{cb}$ is possible
through the leptonic
decay $B_c\rightarrow \tau\nu$, which has decay rate
\begin{equation}
\Gamma(B_c\rightarrow \tau\nu) =
\frac{G_F^2}{8\pi}|V_{cb}|^2f_{B_c}^2m_{B_c}^3\frac{m_\tau}{m_{B_c}^2}
\left(1-\frac{m_\tau}{m_{B_c}^2}\right)^2,
\end{equation}
where the decay constant is defined through
\begin{equation}
\left\langle 0 |A_0 |B_{(c)}\right\rangle = f_{B_{(c)}}M_{B_{(c)}}.
\end{equation}
The experimental difficulties
associated with reconstructing the $\tau$ lepton from its decay products are considerable
\cite{chiladze99} and this decay is yet to be observed experimentally. This, however, offers a
unique possibility for lattice theorists: the prediction
of the leptonic decay rate from lattice computations. Successful predictions lend significantly
more weight to our belief in the validity of lattice results than postdictions and are an important
aspect of testing lattice QCD. 

Lattice determinations of the semileptonic $B\rightarrow D$ decays have so far been dominated by
calculations from the Fermilab
Lattice/MILC collaboration, which has studied various semileptonic
$B$ decays, including $B\rightarrow D\tau\nu$ \cite{bailey12a}, $B\rightarrow D^\ast\ell\nu$
\cite{bernard09} and the ratio of $\overline{B}\rightarrow
D^+\ell^-\overline{\nu}$ to $\overline{B}_s\rightarrow D^+_s\ell^-\overline{\nu}$
\cite{bailey12b}. Calculations of both the semileptonic $B_{(s)}\rightarrow D_{(s)}$ and leptonic
$B_c$ decays are underway by the HPQCD collaboration.

Just as in the heavy-light sector, the role of LPT has been to determine matching coefficients
for quark currents. The Fermilab Lattice/MILC collaboration's results for $B\rightarrow D$
processes required the LPT heavy-heavy current matching
calculations of \cite{elkhadra07}. The HPQCD collaboration is currently undertaking a
 determination of the matching coefficients required for
the semileptonic $B_{(s)}\rightarrow D_{(s)}$ and leptonic $B_c$ meson processes \cite{monahan12}.

\subsection{$B$ mixing}\label{sec:bmix}

The phenomenon of meson-antimeson oscillation is of particular interest to
BSM physics hunters. Meson-antimeson mixing
is a FCNC process and so, in the Standard Model, proceeds through
one-loop box diagrams. This makes meson-antimeson mixing a particularly
powerful contraint on both CKM unitarity and possible BSM physics.

Meson-antimeson mixing occurs in both the $B_d^0$ and $B_s^0$ systems and
is primarily mediated by top quarks. Determinations
of the oscillation frequencies, $\Delta M$ and $\Delta M_s$ respectively,
directly constrain the CKM unitarity triangle through the
Standard Model relations
\begin{equation}
\Delta M_q \propto |V_{tq}^\ast V_{tb}|^2f_{B_q}^2\widehat{B}_{B_q},
\end{equation}
where $q$ may be either a $d$ or $s$ quark. Here the $f_{B_q}$ are the
decay constants discussed in the previous section and the
$\widehat{B}_{B_q}$ are the renormalization group invariant bag
parameters, The decay constants and bag parameters parameterize the
nonperturbative matrix elements defined
by
\begin{equation}\label{eq:bmixmatrix}
\left\langle \overline{B}_q^0\Big|
\Big(\overline{q}\gamma^\mu(1-\gamma_5)b\Big)
\Big(\overline{q}\gamma^\mu(1-\gamma_5)b\Big)
\Big|B^0_q\right\rangle = \frac{8}{3}M_{B_q}^2f_{B_q}^2B_{B_q},
\end{equation}
which are expressed in terms of the scale dependent bag parameters
$B_{B_q}$. The relation between $B_{B_q}$ and $\widehat{B}_{B_q}$
is known perturbatively to next-to-leading order \cite{ciuchini98,buras00}.

Although both $\Delta M$ and $\Delta M_s$ directly constrain the apex of
the CKM unitarity triangle, the SU(3) breaking ratio $\xi =
f_{B_s}\sqrt{\widehat{B}_{B_s}}/f_B\sqrt{\widehat{B}_{B_d}}$ provides
a more stringent bound on CKM matrix unitarity because many systematic
uncertainties cancel in this ratio. New physics may couple to the $B^0$ and $B_s^0$
systems differently and so separate constraints from
both $\Delta M$ and $\Delta M_s$ are required.

Precise determinations of the matrix elements in eq.~\eqref{eq:bmixmatrix} require lattice QCD. And
extracting continuum
results from these lattice computations is once again a job for LPT.

An exploratory study by the RBC/UKQCD collaboration found $\xi = 1.13(12)$ using a single lattice
spacing on $n_f = 2+1$ dynamical domain wall configurations with static $b$ quarks
\cite{albertus10}. LPT was used to improve the axial-vector current and match the results to the
continuum \cite{loktik07,christ07b}. A more precise calculation by the HPQCD collaboration employed
AsqTad light valence quarks and NRQCD $b$ quarks on ensembles with $n_f = 2+1$ AsqTad sea quarks
\cite{gamiz09}. Using the perturbative matching results of \cite{gamiz08}, the HPQCD collaboration
obtained $\xi = 1.258(33)$. These results are currently being updated using the HISQ action for the
light valence quarks, which should reduce the uncertainty. Most recently, the Fermilab Lattice/MILC
collaboration found $\xi = 1.268(63)$ using the same $n_f=2+1$ AsqTad configurations, with AsqTad
light valence quarks and $b$ quarks implemented with the Fermilab action. Details of the
perturbative calculation required for this result are forthcoming.

\subsection{$b$ quark mass determinations}\label{sec:bqmass}

Although not directly measurable, the mass of the $b$ quark is nevertheless an important object of study, both 
as a free parameter of the Standard Model and as an input into searches for BSM physics.

The most precise nonperturbative result currently available for the $b$ quark mass (expressed in
the modified minimal subtraction, or $\msbar$, scheme) is $m_{\MS}(m_{\MS},n_f=5) = 4.165(23)$
GeV, from a relativistic HISQ calculation by the HPQCD collaboration \cite{mcneile10}. Two
different determinations using $n_f = 2$ ensembles from the ALPHA and
ETM collaborations are in agreement with the results from heavy HISQ
simulations. Using nonperturbatively tuned HQET, the ALPHA collaboration obtained a preliminary
result of
$m_{\MS}(m_{\MS}) = 4.22(10)(4)_{\text{z}} $ GeV
\cite{bernardoni12}, where the first error includes statistical and systematic uncertainties and the second arises
from the quark mass renormalization. An alternative approach
was taken by the ETM collaboration, using two different methods to find a (preliminary)
average result of $m_{\MS}(m_{\MS}) = 4.29(13)_{\text{stat}}(4)_{\text{sys}}$ GeV
\cite{dimopoulos12}. Here the first uncertainty is statistical and the second the
total systematic uncertainty.

\subsubsection{A physics example: LPT in the HPQCD approach}

An example that illustrates the importance of LPT for precision $B$ calculations is the
extraction of the $b$ quark mass from lattice NRQCD computations undertaken by the HPQCD
collaboration. An early calculation obtained $m_{\MS}(m_{\MS}) = 4.4(3)$ GeV \cite{gray05a},
where the uncertainty was dominated by the one-loop matching calculation needed to extract the
continuum result. In recently completed work \cite{lee12}, the one-loop calculation was extended to
two loops using a mixed approach combining automated LPT and weak coupling simulations to extract
the heavy quark energy shift. Preliminary indications suggest that this will reduce the
uncertainties to the sub percent level: a real vindication of the power of LPT for precision lattice
NRQCD.

\subsection{Heavy meson spectroscopy}\label{sec:bspec}

My focus has so far been the application of LPT to
precision tests of the Standard Model in the hunt for BSM physics. These
efforts require precise theoretical calculations, which, for
nonperturbative QCD, often means lattice computations. But how can we be
sure that the precision of our computations really is improving and that we
are not missing sources of systematic uncertainty?

To assuage these worries and to ensure that we really are
improving the precision of our lattice calculations we need a
proven testing ground for our results. In the $B$ sector, this testing ground is heavy meson
spectroscopy.

The heavy
meson spectrum is an excellent arena for testing and quantifying uncertainties
associated with lattice calculations \cite{davies04}. There are three reasons for this:
firstly there are a number of gold-plated states that are experimentally
and theoretically well defined; secondly there are multiple states that
can be used as parameter-free tests of the current status of lattice
calculations and their associated uncertainties; and thirdly and most
importantly, there remain a number of experimentally undiscovered states
that allow lattice QCD predictions to be tested. Predictions have
considerably more credibility that postdictions and provide the most
stringent tests of lattice computations.

An early vindication of the methods of precision lattice QCD occurred in the ``prediction'' of the
$B_c$ meson mass. Although this mass was
reconstructed with a precision of about 400 MeV by the CDF collaboration in 1998 \cite{cdf98}, the
resolution was sufficiently poor that a later lattice calculation could claim a ``prediction'' of
$m_{B_c} = 6304\pm12^{+18}_{− 0}$ MeV \cite{allison05}. This result was confirmed by
observations at the Tevatron by the CDF collaboration \cite{abulencia06} and the current world
average value is $m_{B_c} = 6.277\pm0.006$ MeV \cite{pdg12}. The HPQCD recently updated its result
to $6.280\pm10$ MeV \cite{gregory11}, in excellent agreement with the experimental value.

\begin{figure}[h]\label{fig:bspec}
\centerline{\psfig{file=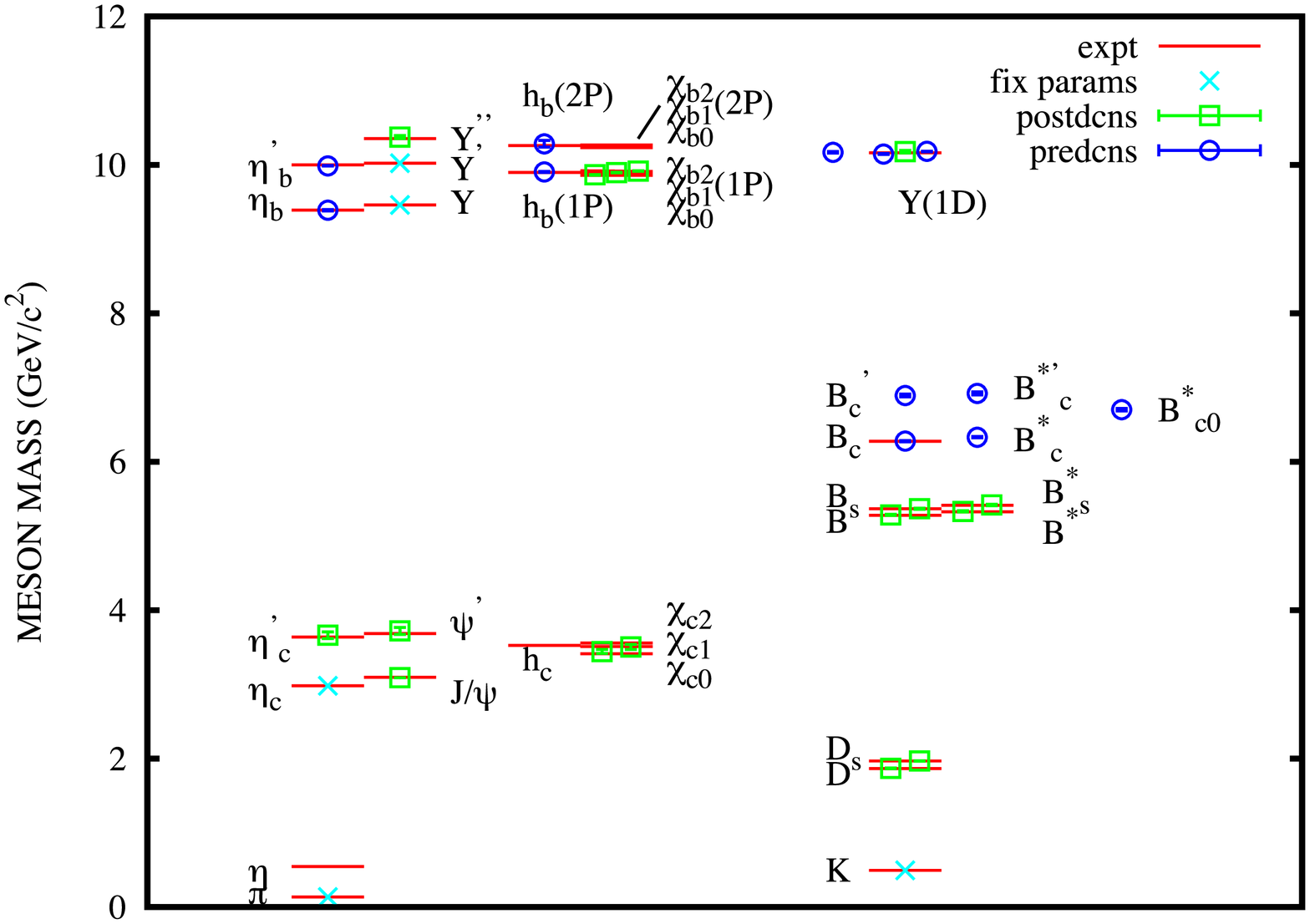,width=0.8\textwidth}}
\vspace*{8pt}
\caption{The current status of precision $B$ spectroscopy by the HPQCD
collaboration. Figure reproduced courtesy of R.J.~Dowdall \emph{et al.}~$^{14}$.}
\end{figure}

Heavy quark spectroscopy remains an important proving ground for lattice computations. The HPQCD
collaboration has pioneered precision B spectroscopy using the HISQ action for
light valence quarks and NRQCD $b$ quarks. This programme spans accurate postdictions for both
heavy-light and heavy-heavy systems \cite{dowdall12a,gregory11} and precise predictions, such as the
mass of the vector $B_c^\ast$ meson \cite{gregory10}, the spectra for excited $B_c$ states
\cite{dowdall12b} or the $D$-wave spectra of the $\Upsilon$ \cite{daldrop11}. The current status of
the $B$ spectrum from the HPQCD collaboration is illustrated in Fig.~\ref{fig:bspec}.

The Fermilab
Lattice/MILC collaboration has addressed heavy quark spectroscopy in \cite{bernard11,burch10},
whilst the
RBC/UKQCD collaboration has used relativistic heavy quarks to examine the low-lying heavy-heavy
bound states \cite{aoki12}. The ETM and ALPHA collaborations are also currently
undertaking
calculations to address $B$ spectroscopy. For heavy-heavy systems the spectrum and
radiative decays of excited $\Upsilon$ states have recently been computed \cite{lewis11,lewis12a,lewis12b}.
 Work has also been carried out on triply-heavy $b$ baryon
spectroscopy, including not only the ground states \cite{meinel10a,meinel10b}, but even the excited
states \cite{meinel12}. 

\subsubsection{A physics example: LPT in the HPQCD approach}

In the context of precision heavy meson spectroscopy, LPT has been primarily used to
improve heavy quarks actions. The importance of this apparently small role is well
illustrated by recent work on the hyperfine splitting of the
$\Upsilon$ meson from lattice NRQCD \cite{dowdall12a}. In early determinations of this
splitting, the HPQCD collaboration used a tree-level value of the coefficient of the chromomagnetic
interaction in the NRQCD action \cite{gray05a}. The results were
not in good agreement with experimental data. A perturbative calculation of the one-loop correction
to the chromo-magnetic interaction in NRQCD brought the lattice results in line
with those from experiment \cite{hammant11}. This calculation
demonstrates the effectiveness of the Symanzik improvement programme for lattice NRQCD
and perfectly illustrates the importance of LPT to precision $B$ physics.

\section{Conclusions}\label{sec:conc}

The past decade has heralded a range of impressive successes for $B$ physics on the lattice.
Precise predictions and accurate postdictions, particularly in $B$ spectroscopy, have convincingly
demonstrated the effectiveness of contemporary lattice calculations. Improving determinations of
decay constants and form factors have facilitated ever-tightening constraints on the unitarity of
the CKM
matrix and placed more stringent bounds on BSM physics effects in rare $B$ decays.

As more experimental data arrive,
particularly from the LHC, further improvements and greater precision will be required from
lattice computations. A close look at recent lattice QCD calculations
reveals that lattice perturbation theory has been crucial in the development of
precise theoretical results. A spectrum of applications of lattice perturbation theory can be found
in contemporary lattice calculations, from the explicit --- matching calculations used to
extract continuum results --- to the more implicit --- in constructing, developing or
improving lattice actions. Lattice perturbation theory
has a home in the work of all the major lattice collaborations and will
continue to help clarify our understanding of $B$ physics.

\appendix

\section*{Acknowledgments}

I would like to thank Dirk Hesse and Christoph Lehner for useful discussions and helpful
corrections regarding their automated LPT routines. Particular
thanks go to Matthew Wingate for reading an early draft of the manuscript. This work has been
partially supported by U.S. DOE Grant No.~DE-FG02-04ER41302.

\end{document}